\documentclass{aa} %%
\usepackage[T1]{fontenc}
\usepackage[varg]{txfonts}
\usepackage{graphicx}
\usepackage{txfonts}
\usepackage{natbib}
\usepackage{verbatim}
\bibpunct{(}{)}{;}{a}{}{,}
\title{Time variations of the narrow \ion{Fe}{ii} and \ion{H}{i} spectral emission lines from the close 
vicinity of $\eta$ Carinae during the spectral event of 2003 \thanks{Based on data collected at the Pico dos Dias Observatory (LNA-MCT)}}
\author{H. Hartman \inst{1} \and A. Damineli \inst{2} \and S. Johansson\inst{1} \and V.S. Letokhov\inst{3,1}}
\institute{Lund Observatory, Lund University, P.O. Box 43, S-22100, Lund, SWEDEN 
\and Instituto de Astronomia, Geoffisica e Ciencias Atmosfericas, University of Sao Paolo, 
Rua do Matao 1226, 05508-900 Sao Paulo, BRAZIL \and Institute of Spectroscopy, Russian Academy of Sciences,
Troitsk, Moscow region, 142190, RUSSIA}
\date{Received <date> / Accepted <date>}
\abstract{
The spectrum of \object{Eta Carinae} and its ejecta shows slow variations over a period of 5.5 years. 
However, the spectrum changes drastically on a time scale of days once every period called the 'spectral event'. 
We report on variations in the narrow emission line spectrum of gas 
condensations (the Weigelt blobs) close to the central star during a spectral 
event. The rapid changes in the stellar radiation field illuminating the blobs 
make the blobs a natural astrophysical laboratory to study atomic 
photoprocesses. The different responses of the HI~Paschen lines, fluorescent 
$<$\ion{Fe}{ii}$>$ lines and forbidden [\ion{Fe}{ii}] lines allow us to identify 
the processes and estimate physical conditions in the blobs.
This paper is based on observations from the Pico dos Dias Observatory (LNA/Brazil) during the previous event in June 2003. 
}
\titlerunning{Time variations of FeII and HI lines during the spectral event of 2003}
\authorrunning{Hartman et al.}
\begin{document}
\maketitle
\section{Introduction}
A general description of Eta Carinae and its nebulosities is given by 
Davidson and Humphreys (1997). During the last decade more detailed information
about the complex object has been obtained thanks to high-resolution spatial and spectral
observations performed with the Hubble Space Telescope \citep{GID01} and the VLT/UVES 
instrument \citep{WSB05,SWB05}. The observational data have shown that Eta 
Carinae and its close environment contain a number of unique spectroscopic challenges 
formulated by \citet{D01}. 

Eta Carinae is one of the most luminous and massive stars in the Galaxy, and the 
site of huge non-thermal stellar eruptions as well as the source of radiation 
with very different properties. Wide Doppler-shifted spectral lines originate 
from fast moving ejecta in the Homunculus, thereby being a source of information about 
the kinematics of the expanding asymmetrical hollow lobes \citep{SMG04}. Similar information
is obtained from several velocity systems detected by numerous blue-shifted absorption lines 
of many elements \citep{GVB05}.
A forest of narrow spectral emission lines originate from dense slowly-moving ejecta near Eta Carinae \citep{DEW95}, 
which were detected and resolved as a few objects (the Weigelt blobs B, 
C. D) by Weigelt and Ebersberger (1986) using speckle-interferometry. A two-zone photoionization model of 
the Weigelt blobs was developed by Johansson and Letokhov in 2001-2004, which provides 
a natural explanation of the origin of the narrow emission lines from a HI zone and the wider 
lines from a HII zone inside the blobs. This photoionization model is also quite 
useful for the explanation of a time variation observed in the narrow FeII and HI spectral 
lines during the spectroscopic event in 2003, which is subject of the present paper. 
The densities and temperatures used in our model of the blobs are similar to those derived by
\citet{VGB02}.

The neighborhood of the massive star Eta Carinae is for many reasons a giant natural space laboratory for 
atomic astrophysics. Firstly, in the immediate vicinity of the 
central source we find the Weigelt blobs B, C, D, having a hydrogen concentration of the order of 
%%Weigelt and Ebersberger (1986) \citet{WE86} discovered several gas condensations (Weigelt blobs B, C, D) 
10$^8$ cm$^{-3}$ \citep{DH97}. 
%%(Davidson and Humphreys, 1997). 
At such a low hydrogen concentration, collisional processes occur on a time scale much larger than 1~s. 
For this reason, the radiative relaxation of excited atoms and ions occurs under 
collision-free conditions, which somewhat facilitates the interpretation of the complicated observed spectra.

Secondly, the Weigelt blobs are located at angular distances of the order of 
0.1-0.2 arcsec from the central star, i.e., at around 300 stellar radii ($r_\mathrm{s}$),
where $r_\mathrm{s}$ = 3$\cdot10^{13}$ cm \citep{BKS03}. %%(Boekel et al., 2003).
The small distance provides for 
a strong photoionization of the blobs and the stellar wind by the Lyman continuum 
radiation, Ly$_\mathrm{c}$ from the photosphere of Eta Carinae. A substantial proportion of the absorbed 
radiation is converted into H Ly$\alpha$ radiation by photorecombination. Thus, the 
close ejecta of Eta Carinae are irradiated by intense Ly$\alpha$ radiation corresponding to 
an effective spectral temperature above 10\,000~K \citep{JL04a}. %%(Johansson and Letokhov, 2004a).

Thirdly, the coincidence between the wide H Ly$\alpha$ spectral line and a 
dozen of \ion{Fe}{ii} absorption lines \citep{BJW79,JJ84}
%%(Brown et al., 1979; Johansson and Jordan, 1984)
 gives rise to a number of fluorescence <\ion{Fe}{ii}> lines from highly 
excited states. Especially conspicuous among them are the abnormally bright 
2507/09-Å UV lines of \ion{Fe}{ii} \citep{JH93}. %%(Johansson and Hamann, 1993).
 The 
fluorescence lines have a small spectral width, since they originate in the 
relatively cold HI regions of the partially ionized Weigelt blobs \citep{JZ99,JL01e}.
%%(Johansson and Zethson, 1999; Johansson and Letokhov, 2001). 
The first indication that the narrow \ion{Fe}{ii} lines 
actually are formed in the Weigelt blobs was based on observations with the FOS instrument
onboard the Hubble Space Telescope \citep{DEW95} %%(Davidson et al., 1995) 
and later confirmed by numerous observations of Eta Carinae with GHRS and STIS 
\citep{DEJ97,GID99}.
%%(Davidson et al., 1997; Hamann et al., 1999).

In the fourth place, when attempting to explain the origin of the abnormally 
bright 2507/09~\AA\ lines from a detailed analysis of 
the \ion{Fe}{ii} spectrum and the radiative relaxation pathways of high-lying, 
H Ly$\alpha$-pumped \ion{Fe}{ii} levels, the idea of stimulated 
emission of radiation (the astrophysical laser) in transitions from long-lived 
(pseudo-metastable) \ion{Fe}{ii} states of has been proposed \citep{JL02a}. %%(Johansson and Letokhov, 2002).
 The stimulated emission yields a purely radiative cycle 
of \ion{Fe}{ii} in a four-level scheme pumped by intense 
H Ly$\alpha$ radiation. The close cycle gives rise to the intense UV spontaneous emission lines of 
\ion{Fe}{ii} \citep{JL03,JL04d}. %%(Johansson and Letokhov, 2003; 2004b).

Finally, as a fifth point, the discovery by \citet{D96} %%Damineli (1996) 
of the periodic (5.5 years) 
reduction of the intensity of photospheric He lines  for 
a few months, called "the spectroscopic event", offers a unique possibility to use time probing 
of the photophysical processes taking place in the blobs.  
The periodicity led \citet{DCL97} %%Damineli, Conti and Lopes (1997) 
to propose 
a binary system composed by evolved massive stars. The expected 
wind-wind collision is revealed by an X-ray source, which is variable and whose hard
X-rays is deeply absorbed \citep{PC02}. %%(Pittard and Corcoran 2002). 
The detection of phase locked \ion{He}{II} 4686~\AA\ emission is an 
indirect evidence that the secondary star is much hotter 
than the primary and is the source of ionization for
the high excitation lines seen in the Weigelt blobs
\citep{SD04}. %%(Steiner and Damineli 2004).
It is especially 
valuable that the relatively small distance (7500 l.y.) to Eta Carinae  makes it 
possible to observe the spectral response of the blobs to such a periodic change 
in the intensity of the photoionizing radiation from Eta Carinae with a high 
spatial (angular) and spectral resolution by means of the $HST$/STIS facility 
\citep{GID01}. %%(Gull et al., 2001).  

This paper presents the result of ground-based spectroscopic observations of narrow 
\ion{Fe}{ii} and H Paschen lines from the Weigelt blobs of 
Eta Carinae made during the spectral event of 2003. We selected these 
\ion{Fe}{ii} lines, which are narrow relative to stellar wind lines, and emitted by the cool (HI) zone, 
and also narrow Paschen lines from the hot HII zone of the slow-moving blobs. First (Sect. 2) 
we briefly describe the basic photoprocesses occurring in the Weigelt blobs 
receiving radiation from the central source, a primary and a secondary star, and also the 
intense Ly$\alpha$ radiation from the stellar wind. The observations 
are described in Sect.~3. In the subsequent sections, we describe and interpret 
the temporal behavior of the intensity of: the Paschen lines  
(Sect. 4), the fluorescent  <\ion{Fe}{ii}> lines, including the 
\ion{Fe}{ii} fluorescent probing of the width of the wind component of Ly$\alpha$ (Sect. 5),
 and forbidden [\ion{Fe}{ii}] lines from low-lying metastable states (Sect. 6). In 
the conclusion (Sect. 7), we discuss the validity of our simple model for the basic 
photoprocesses taking place in the Weigelt blobs (Sect. 2) on the basis of the 
observational data presented.

\section{Basic Photoprocesses Occurring in the Weigelt Blobs}

Before discussing the observational data available, it seems worthwhile to 
present a general picture of the photoprocesses taking place in the Weigelt 
blobs. This picture has emerged during the course of investigations that are aimed at 
explaining the abnormally bright 2507/09~\AA\ spectral lines of \ion{Fe}{ii} 
\citep{JL02a,JL04a,JL04d}. %%(Johansson and Letokhov, 2002; 2004a, b).

\begin{figure*} %%Figure 1
\sidecaption
\resizebox{12cm}{!}{\rotatebox{00}{\includegraphics{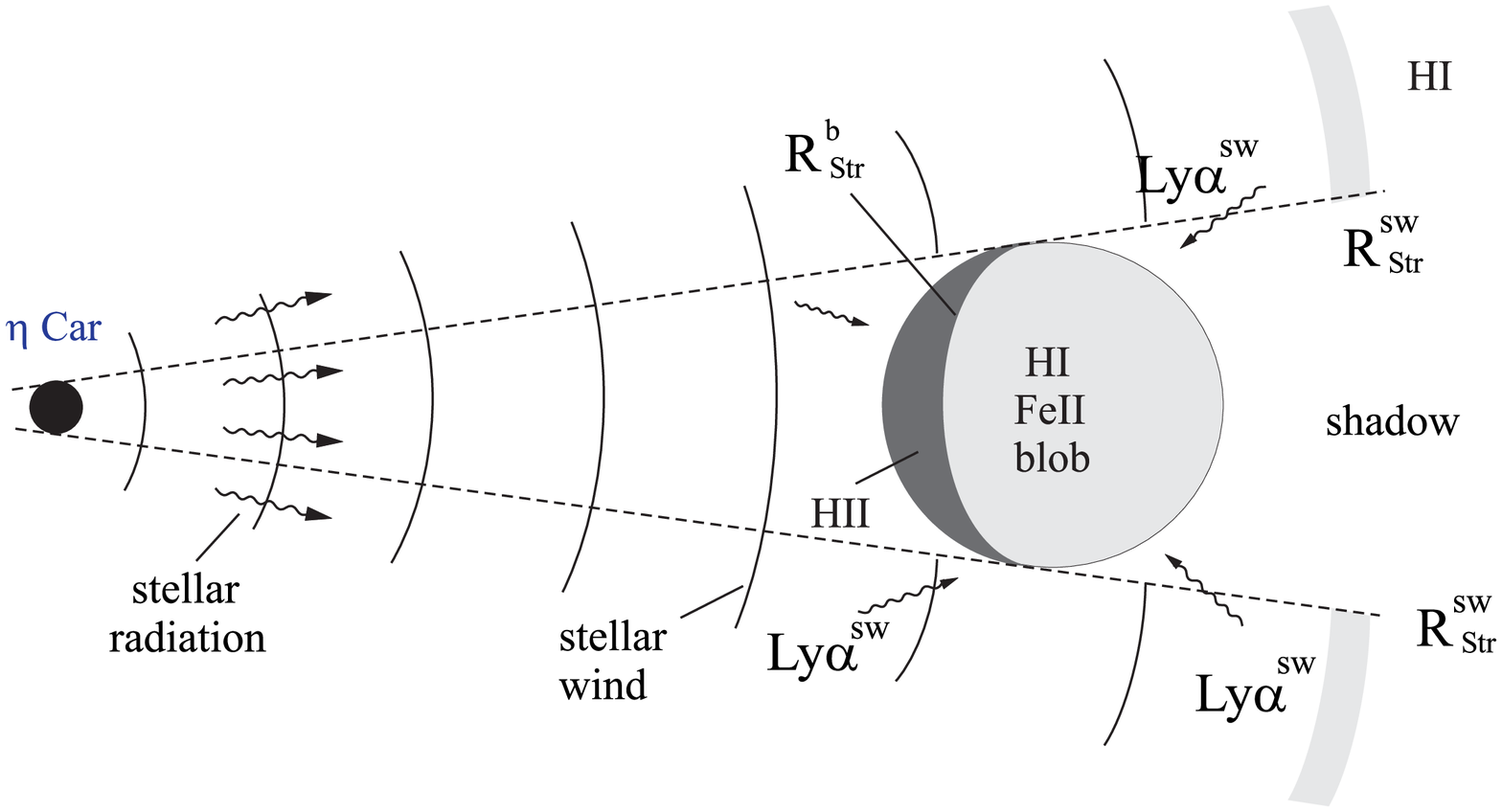}}} %%rr1v2.eps
\caption{A sketch of the formation of fluorescent and forbidden spectral lines in a Weigelt blob with the \ion{H}{ii} and \ion{H}{i} zones irradiated by Ly$\alpha^{\mathrm{sw}}$ from the stellar wind ($R_{\mathrm{Str}}^{\mathrm{b}}$ and $R_{\mathrm{Str}}^{\mathrm{sw}}$ are the Strömgren boundaries in the blob and stellar wind, respectively).}
\end{figure*}
 
Figure 1 presents a simplified 
geometry of an average blob (blob B, for example) surrounded by the stellar wind of Eta 
Carinae. Consider the 
case where the Strömgren boundary for the stellar wind, 
$R_{\mathrm{Str}}^{\mathrm{sw}}$, is located behind the blob, i.e., 
$R_{\mathrm{Str}}^{\mathrm{sw}}>R_\mathrm{b}$ , where $R_\mathrm{b}$ is the 
distance between the blob and the central source. In that case, the ionizing radiation 
from the central source reaches the surface of the blob and photoionizes its front 
part. Thus, the Strömgren boundary for the blob, $R_{\mathrm{Str}}^{\mathrm{b}}$, 
intersects the blob, dividing it into a hot (HII) and a cold (HI) zone. The 
Ly$\alpha$ radiation generated in the HII region of the blob and in the stellar 
wind irradiates the blob on all sides. The Ly$\alpha$ radiation from the stellar 
wind has a broad spectrum determined by the terminal velocity $v_\mathrm{t} 
\simeq$~650km/s: $\Delta \nu_{\mathrm{sw}}=2(v_\mathrm{t}/c) \nu_0$, where 
$\nu_0$ is the central frequency of the Ly$\alpha$ radiation. The spectrum of 
the Ly$\alpha$ radiation incident on the slow-moving blob from the stellar wind 
can be shifted either toward the blue or toward the red, depending on the 
direction of the radiation: when the blob is irradiated from behind (in opposite 
direction of its movement), the radiation spectrum is red-shifted, and when it is 
irradiated in the same direction as it moves, the spectrum is blue-shifted (Fig. 
1 and 2).

\begin{figure} %%Figure 2
\resizebox{8.8cm}{!}{\includegraphics{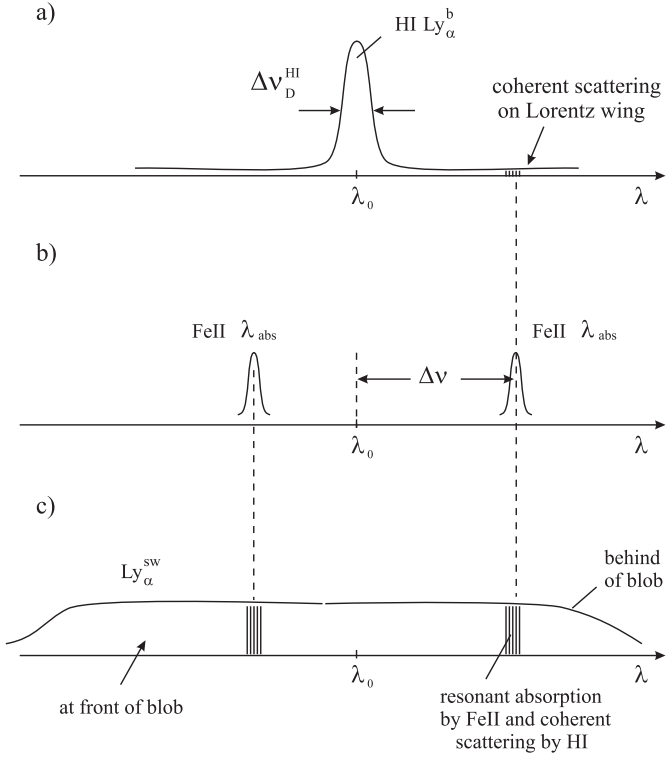}}
\caption{Illustration of the mutual positions of spectral lines: (a) Ly$\alpha^{\mathrm{b}}$ in the blob; 
(b) \ion{Fe}{ii} absorption lines; (c) wide Ly$\alpha^{\mathrm{sw}}$ from the stellar wind.}
\end{figure}

The Fe$^+$ ions are formed both in the hot (HII) and in the cold (HI) 
zones of the blob, since the optical density of the Fe I-ionizing radiation 
(919 Å$< \lambda_{\mathrm{ion}} < $1520 Å) is not very high. In other words, the Fe I 
photoionization rate is much higher than the \ion{Fe}{ii}-electron recombination rate,
$W_{\mathrm{ph}}>>W_{\mathrm{rec}}$. In fact, the 
mean FeI photoionization cross section $\left< \sigma_{\mathrm{ph}} \right> \simeq 10^{-
18} \mbox{ cm}^2$, whereas the \ion{Fe}{ii} recombination coefficient 
$\alpha_{\mathrm{rec}} \simeq 10^{-11} \mbox{ cm}^3 \cdot \mathrm{s}^{-1}$ 
\citep{BP98}. %%(Bautista and Pradhan, 1998). 
With an effective temperature $T \simeq 30\,000$~K 
of the ionizing radiation from the central source, the ionizing radiation flux 
in the frequency region corresponding to the 5~eV gap between 
the ionization potentials of FeI and HI is $\Phi_{\mathrm{phr}} \simeq 
5\cdot10^{23}$ photons/cm$^2 \cdot$s$\cdot$sr. Therefore, the photoionization 
rate of Fe I is
\begin{equation} 
W_{\mathrm{ph}}=<\sigma_{\mathrm{ph}}> \Phi_{\mathrm{ph}} \Omega \simeq 0.5-5 
\hspace {2mm} {\mathrm s}^{-1}, 
\end{equation}                                          
where $\Omega = \frac{1}{4}(r_{\mathrm{s}}/R_{\mathrm{b}})^2 \simeq 10^{-6}-
10^{-5}$ is the dilution factor, and $R_{\mathrm{b}}$ is the distance between the blob 
and the central source. At the same time, the \ion{Fe}{ii} + e recombination rate 
is
\begin{equation} 
W_{\mathrm{rec}} \simeq \alpha_{\mathrm{rec}} \cdot n_{\mathrm{e}} \simeq 10^{-
8}-10^{-7} \mathrm{s}^{-1} \approx 10^{-3} \mbox{ to } 10^{-2} \mbox{ day}^{-1},
\end{equation}
where it is assumed that the electron density in the HI zone is governed mainly 
by the photoionization of FeI, i.e. $n_\mathrm{e} \simeq 
N(\mathrm{\ion{Fe}{ii}})= 10^{-4} N_{\mathrm{H}} \simeq 10^3-10^4 \mbox{ cm}^{-
3}$ for $N_{\mathrm{H}} \simeq 10^7-10^8 \mbox{ cm}^{-3}$.

The Ly$\alpha$ radiation responsible for the photoselective excitation of \ion{Fe}{ii}
can stem from two sources. First, an intense Ly$\alpha$ radiation generated in the HII zone 
of the blob enters its HI zone. However, the penetration depth of this 
spectrally narrow radiation is restricted by the high optical density of 
absorption in the Ly$\alpha$ transition, and the significant ($\pm$2.5 Å) 
detuning of the \ion{Fe}{ii} absorption lines relative to Ly$\alpha$. This 
matter was considered by \citet{KJL02}. %%(Klimov et al., 2002). 
The 
situation is more clear when one consider the Ly$\alpha$ radiation 
generated in the stellar wind in the geometry shown in Fig. 1. This radiation 
has a broad spectrum for the slow-moving HI and \ion{Fe}{ii} zones in the blob. The 
\ion{Fe}{ii} ion can undergo photoselective excitation when it is in exact 
resonance with the red-shifted component of the Ly$\alpha$ radiation from the 
stellar wind (Fig. 2).

\begin{figure} %%Figure 3
\resizebox{8.8cm}{!}{\includegraphics{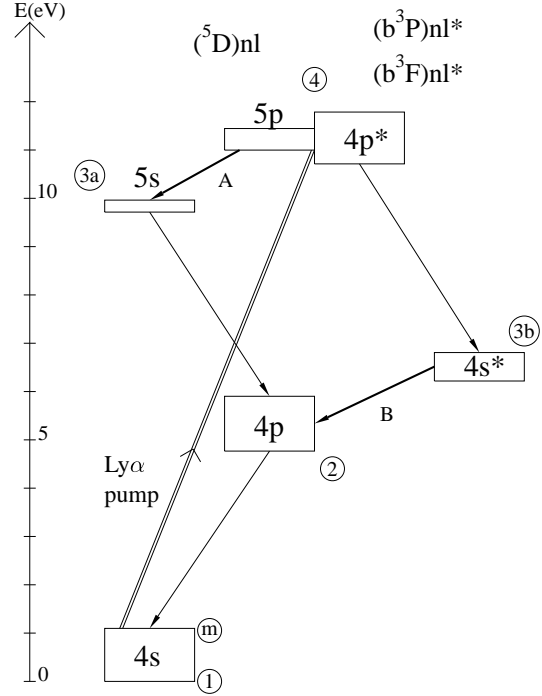}}
\caption{Partial energy level diagram of \ion{Fe}{ii}, including only the set of levels discussed in the present paper. The lines of relevance are marked with A and B. These are primary and secondary decays, respectively, of the 
H~Ly$\alpha$ pumped levels. The circled numbers are introduced to make the discussion in the text easier to follow.}
\end{figure}

Next question concerns the optical depth in the \ion{Fe}{ii} absorption 
line from a low-lying metastable state "$m$" (see Fig. 3) to the high-lying state 4 
that leads to a chain of fluorescence transitions  
4 $\rightarrow$ 3 $\rightarrow$ 2 $\rightarrow$ ...  and so on. The optical 
depth $\tau_{m4}$ (\ion{Fe}{ii}) can be estimated by the 
standard expression
\begin{equation} 
\tau_{m4}(\mathrm{\ion{Fe}{ii}})=\sigma_{m4}N_m^{\mathrm{\ion{Fe}{ii}}}r_{\mathrm{b}}, 
\end{equation}
where $\sigma_{m4}$ is the absorption cross section in the $m\rightarrow$4 transition,
\begin{equation} 
\sigma_{m4}=\frac{\lambda_{m4}^2}{2\pi}\frac{A_{4m}}{2\pi\Delta\nu_D^{m4}}. 
\end{equation}
$N_m^{\mathrm{\ion{Fe}{ii}}}$ is the population of the low-lying initial state "$m$" 
(one of about 60 metastable states), $r_{\mathrm{b}}$ is the radius of the 
spherical blob, $A_{4m}$ is the Einstein coefficient, and $\Delta \nu_{\mathrm{D}}^{m4}$ 
is the Doppler width of the \ion{Fe}{ii} absorption line. Adopting the following numerical 
data: $A_{4m} \simeq 10^8s^{-1}$, $N_m^{\mathrm{\ion{Fe}{ii}}} \simeq 10^{-2} N(\mathrm{\ion{Fe}{ii}}) 
\simeq 10^{-6} N_{\mathrm{H}}$, $r_{\mathrm{b}} \simeq 0.5\cdot 10^{15}$~cm, 
and $\Delta \nu_{\mathrm{D}}^{m4} = 7\cdot10^{-6} \nu_{\mathrm{0}}$ (for $T\simeq 5000$\,K for HI zone 
\citep{VGB02} %%(Verner et al. 2005) 
we get $\sigma_{m4} \simeq 2.5 \cdot 10^{-14} \mbox{ cm}^2$ and $\tau_{m4} \simeq 1.3 \cdot 10^{-7} N_\mathrm{H}$. 
The magnitude of $N_{\mathrm{H}}$ cannot be less than the critical value 
$N_{\mathrm{cr}}=(0.5\mbox{ to }5) \cdot 10^7 \mbox{ cm}^{-3}$, as it is only in 
this case the cold HI zone of the blob and hence the narrow fluorescence lines 
are formed \citep{JL01e}. %%(Johansson and Letokhov, 2001). 
Therefore, the optical density 
$\tau_{m4}(\mathrm{\ion{Fe}{ii}}) \gtrsim 1-10$. Thus, the Ly$\alpha$ radiation from the 
stellar wind that is in exact resonance with the \ion{Fe}{ii} absorption line ensures a 
preferable excitation of the Fe$^+$ ions in the outer shell of the blob. Even when 
$\tau_{m4} \simeq 10-10^3$, excitation within the limits of 
$\pm 2 \Delta \nu_{\mathrm{D}}^{m4}$ detuning in the wings of the absorption line can also 
provide for the excitation of the Fe$^+$ ions throughout the blob volume. However, 
the intensity of pumping radiation and hence the intensity of fluorescent lines should be 
larger in the outer shell, which is in agreement with observations \citep{SMG04}. %%(Smith et al., 2004).

Those wavelength bands of the stellar-wind Ly$\alpha$ profile, which are in 
resonance with the \ion{Fe}{ii} absorption lines (Fig. 2) are detuned 
substantially from the narrow hydrogen 1s-2p absorption line in the cold zone of 
the blob. However, they are subject to coherent scattering in the Lorentz wings 
of HI, as shown in Fig. 2. The scattering cross section is defined by the 
expression 
\begin{equation} \sigma_{\mathrm{sc}}(\Delta \nu) = \frac{\Delta 
\nu_D^{\mathrm{HI}}\delta\nu_{\mathrm{rad}}}{(2\Delta\nu)^2}\sigma_{\mathrm{sc}}^0,
\end{equation} 
where $\sigma_{\mathrm{sc}}^0$ is the scattering cross 
section at the line center, $\Delta \nu_{\mathrm{D}}^{\mathrm{HI}}$ is the 
Doppler width of HI, $\delta\nu_{\mathrm{rad}}$ is the radiative width of the 
1s-2p transition that determines its Lorentzian profile, and $\Delta \nu$ is the detuning 
of the stellar wind HLy$\alpha$ spectral band, which is in resonance with the 
\ion{Fe}{ii} absorption line. Consider as an example the case where $\Delta \nu 
\simeq 160 \mbox{ cm}^{-1} (\Delta \nu_{\mathrm{D}}^{\mathrm{HI}}=2.4 
\mbox{ cm}^{-1}, \delta\nu_{\mathrm{rad}}=2.5 \cdot 10^{-3} \mbox{ cm}^{-1})$. 
In that case, $\sigma_{\mathrm{sc}}(\Delta \nu)\simeq \sigma_{\mathrm{sc}}^0 
\cdot 6 \cdot 10^{-8}$, which for $\sigma_{\mathrm{sc}}^0=1.4 \cdot 10^{-14} 
\mbox{ cm}^2$ yields  $\sigma_{\mathrm{sc}}(\Delta \nu) \simeq 10^{-21} 
\mbox{ cm}^2$. Hence we have the following estimate for the coherent scattering 
length: $l_{\mathrm{sc}} = 10^{21}$/N$_{\mathrm{H}}$. Again at 
$N_{\mathrm{H}}=(\mbox{0.5 to 5}) \cdot 10^7 \mbox{ cm}^{-3}$ the coherent 
scattering length $l_{\mathrm{sc}} \simeq (\mbox{0.2 to 2}) \cdot 10^{14} 
\mbox{ cm}$, i.e., it is less than the blob radius $r_{\mathrm{b}} \simeq 5 
\cdot 10^{14}$~cm. Thus, the HI zone in the blob is a scattering medium for the 
Ly$\alpha$ photons that selectively excite the Fe$^+$ ions. It takes 
about $N_{\mathrm{sc}} \simeq (r_{\mathrm{b}}/l_{\mathrm{sc}})^2 \simeq 6-600$ 
scattering events for the photons to penetrate inside the blob. This situation 
is more favorable than that for the Ly$\alpha$ photons originating in the 
HII zone of the blob, since they have a more wide Doppler width: $\Delta 
\nu_{\mathrm{D}}^{\mathrm{HII}}\simeq 10$~cm$^{-1}$ and can be absorbed by HI more strongly.

Based on this simple picture of the basic photoprocesses occurring in the blob, 
we can state that the narrow spectral lines observed are formed all over the 
volume of the blob.

\section{Observational Setup and Conditions}

The observations presented in this paper are part of a campaign conducted
 at the Pico dos Dias Observatory (LNA/Brazil) using the Coud\'e focus of 
the 1.6-m telescope. The CCD was a Marconi 2048x4608 pixels and the slit 
width was 1.5 arcsec, resulting in a resolution of R=13\,000 (2 pixels) at 
Balmer~$\alpha$, as measured using a Thorium-Argon lamp. Extractions of the 
spectrum were made in 5 rows along the spatial direction encompassing 
$\sim$2.0 arcsec around the central star. Changing the number of extracted
 rows from 3 to 9 did not produce measurable changes in the line 
intensities. Also changes in the slit width in the range 1-3 arcsec 
produced no significant differences in the spectrum. This is important,
because the seeing was variable along the observing program, especially 
in late July and early August, when the star is at large airmass in the 
beginning of the night. The largest impact on the repeatability of line 
intensities was caused by guiding errors, that occurred at large 
telescope inclinations. For wavelengths longer than 6500Å, we observed 
also the neighbor bright star $\theta$ Car (V=2.6, sp=B0V) in order to 
remove CCD fringing and telluric lines. The stellar continuum was 
normalized by fitting a low order polynomial.

The procedure to measure the equivalent width of the narrow line 
components was integrating the area under the line profile in 
the classical way. However, the narrow components are seated on 
broad components and in some cases, are blended with other lines that 
can be variable. To avoid the contamination of 
variations, we measured the equivalent widths of the narrow
 components referred to the normalized stellar continuum, not 
to the flux at the base of the narrow line components. The 
stellar continuum itself is variable, although only by a small amount 
in the spectral range we are interested.

\begin{figure} %%Figure 4
\resizebox{8.8cm}{!}{\includegraphics{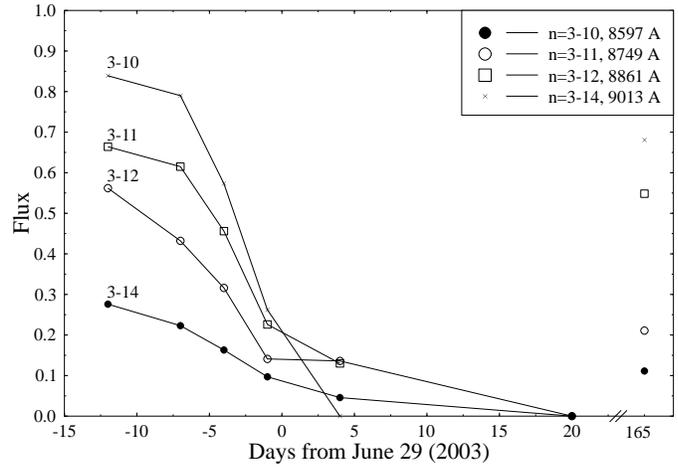}}
\caption{Time behaviour of HI Paschen lines during the spectroscopic event.}
\end{figure}

\section{Intensity Variation of HI Paschen Lines}

In Fig. 4 we present data on the intensity variation of the HI Paschen lines 
(narrow components) from all the Weigelt blobs. The reduction of the intensities 
of the lines can be explained by the reduction of the ionizing radiation flux 
from the central source in Eta Carinae. This can take place if the 
recombination rate W$_{\mathrm{rec}}$ of the H ions in the hot HII zones of 
the Weigelt blobs is higher than the reduction rate of the ionizing radiation 
intensity. Otherwise the intensity of the recombination hydrogen lines 
$n\rightarrow 3$ would remain unchanged during the course of the spectral event.
Note that the cold HI zones of the blobs cannot be the source of the 
Paschen lines. 

From the curves in Fig.~4 we adopt that the intensity of the stellar radiation decreases by a 
factor of 2 during 3 days which gives an estimate of the lower limit of the 
initial (prior to the spectral event) electron density, $n_\mathrm{e}$ in the HII zones 
of the Weigelt blobs:
\begin{equation} 
W_{\mathrm{rec}} = \alpha \cdot n_{\mathrm{e}} \gtrsim \left( 3 \cdot 10^5 \mbox{ s} \right)^{-1}, 
\end{equation}
where $\alpha = \sum \alpha_n = 3 \cdot 10^{-11}/ \sqrt{T_{\mathrm{e}}} 
(\mbox{ cm}^3\mathrm{s}^{-1})$ is the sum of the recombination coefficients for 
each level. From expression (6) we get $n_{\mathrm{e}} \gtrsim 10^7$ cm$^{-3}$ for $T_{\mathrm{e}} \simeq 10^4$\,K in the HII zone. 
This estimate agrees quite well with the estimate of the critical hydrogen 
concentration $N_{\mathrm{cr}}$ in the blobs.

The relative intensities $I(\mathrm{Pa}_n)$ of the HI Paschen lines 
depend on the corresponding recombination coefficients $\alpha_n$ and 
%%radiative transition probabilities $A$(Pa$_n$):
branching fraction from the upper level $BF$(Pa$_n$):
\begin{equation} 
%%I(\mathrm{Pa}_n) \simeq A(\mathrm{Pa}_n) \alpha_n n_{\mathrm{e}}(t) N_{\mathrm{HII}}(t), 
I(\mathrm{Pa}_n) \simeq BF(\mathrm{Pa}_n) \alpha_n n_{\mathrm{e}}(t) N_{\mathrm{HII}}(t), 
\end{equation}
where n$_e$($t$) and $N_{\mathrm{HII}}$($t$) are the time dependencies of the electron 
and ion concentrations during the reduction of the ionizing flux from 
the central source. 
The relative intensities of the Paschen lines in Fig.~4 are in 
qualitative agreement with expression (7). Actually, the diffence in intensities 
of lines "3-$n$" corresponds to the decrease of the recombination coefficient 
$\alpha_n$ with higher quantum number $n$. The temporal decrease of intensities 
for the lines corresponds to the reduction of the electron n$_e$($t$) and proton 
$N_{\mathrm{HII}}$($t$) densities during the spectral event.
%% More detailed conclusions would look too speculative
%%The relative intensities of the Paschen lines in Fig. 4 are in qualitative agreement with expression (7).

\section{Intensity Variations of \ion{Fe}{ii} Fluorescence Lines from Highly Excited States}

\begin{figure} %%Figure 5a,b
\resizebox{8.8cm}{!}{\includegraphics{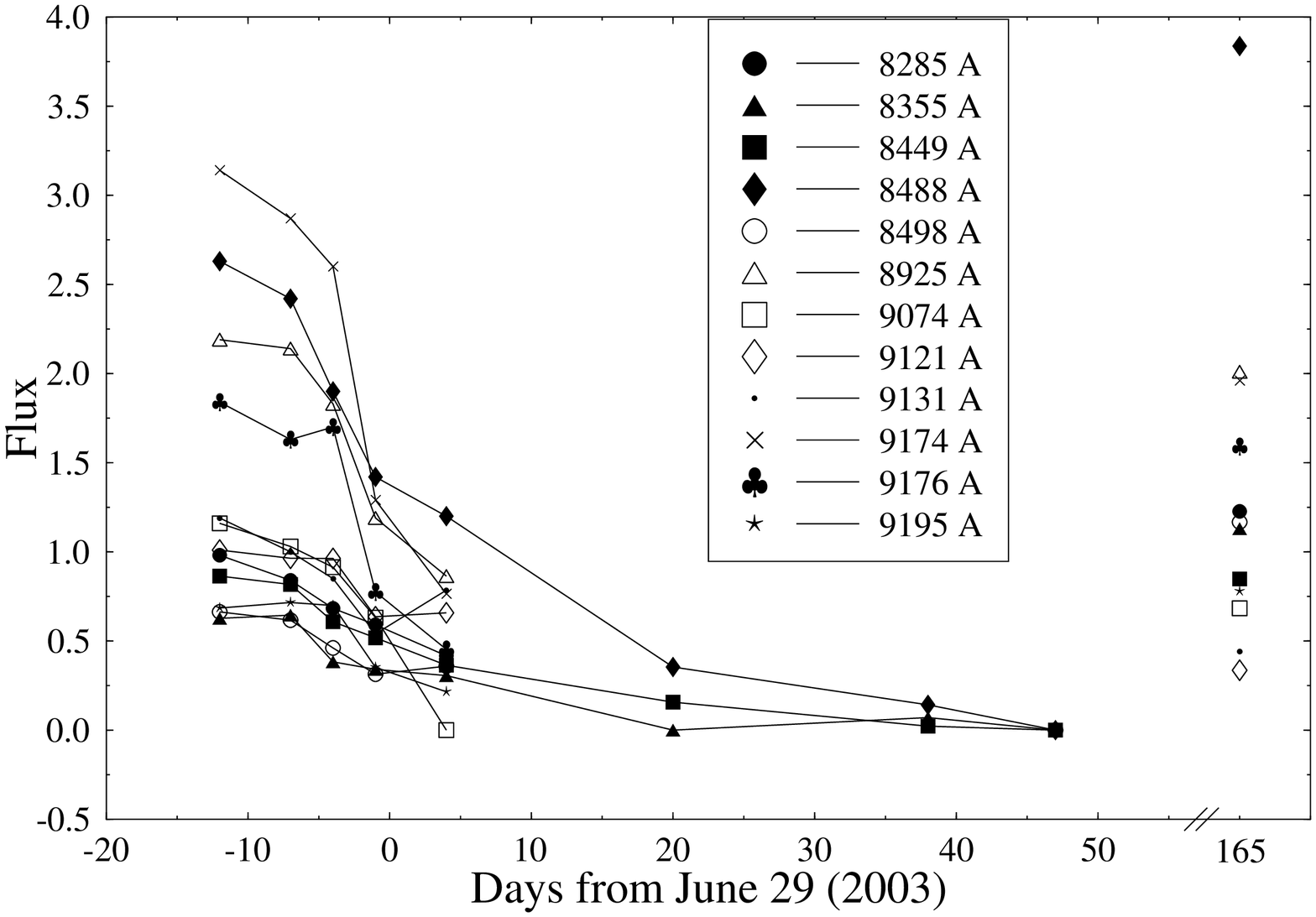}}
\resizebox{8.8cm}{!}{\includegraphics{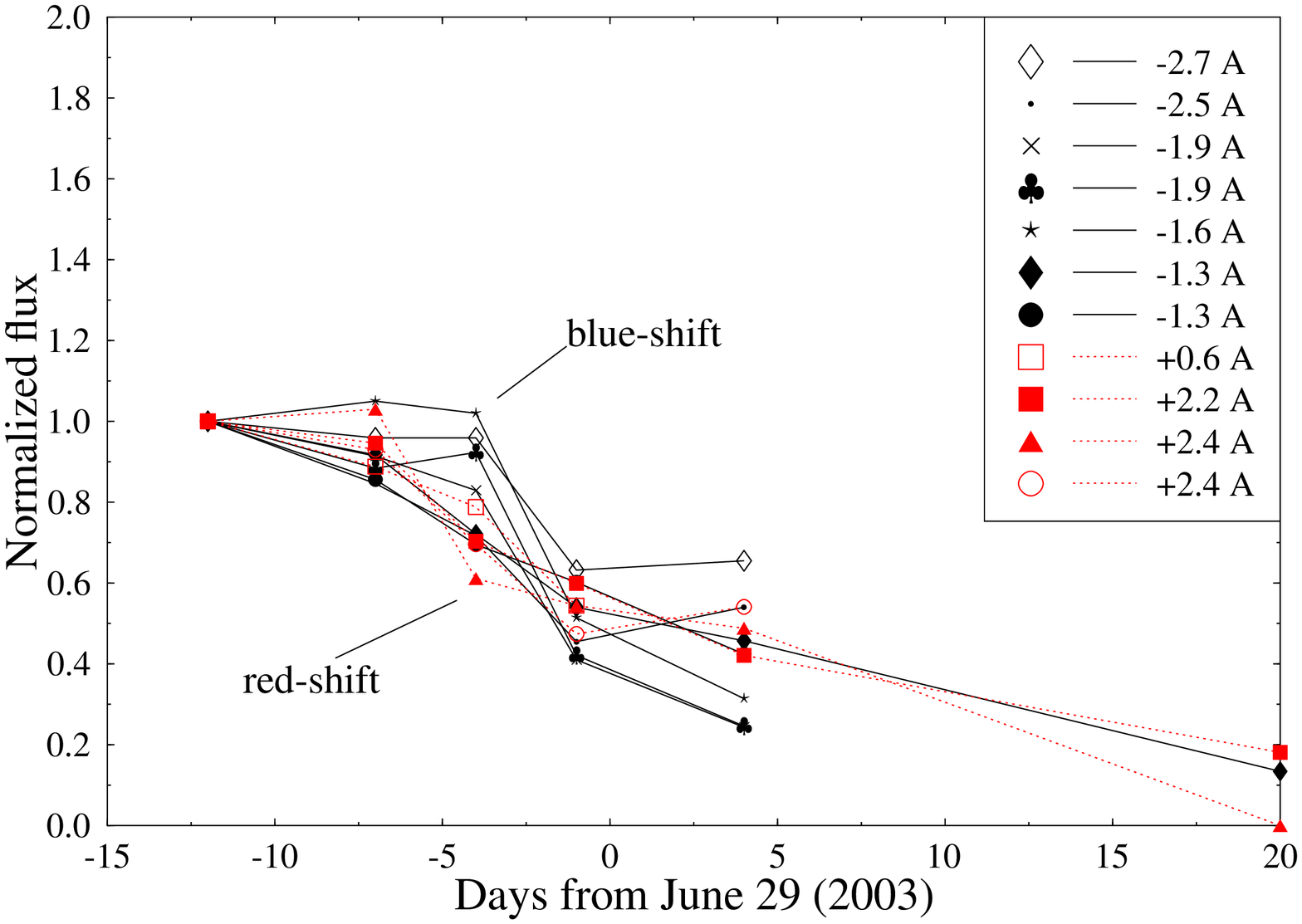}}
\caption{(a) Time behaviour of fluorescent $<$\ion{Fe}{ii}$>$ lines during the spectroscopic event. All 
lines are 5p$\rightarrow$5s transitions (A in Figure 4.) The wavelengths are indicated in the inserted key. All 
lines were not observed after the event. (b) Same as 5a, but normalized to June 17 (-12 days). In the key, 
the wavelength detuning of each excitation channel from HLy$\alpha$ is indicated. A positive value indicates a redshift.
Note the different time scales}
\end{figure}

Figure 5a displays the observed intensity observations for 12 spectral lines 
of \ion{Fe}{ii} excited by Ly$\alpha$ radiation. All these fluorescence lines correspond to the 5p$\rightarrow$5s transition array
(marked A in Fig. 3) in \ion{Fe}{ii}. The rate of intensity reduction of the fluorescence lines is 
approximately the same as that of the recombination Paschen lines (Fig.~4). This 
points to a common origin of the effect of reduction of the intensity of the 
spectral lines of interest. In the case of Paschen lines, the effect is associated 
with the decrease in the intensity of the Lyman continuum radiation, Ly$_\mathrm{c}$, that 
sustains the ionization balance of the \ion{H}{ii} ions in the blobs, while in 
the case of $<$\ion{Fe}{ii}$>$ lines, the effect is due to the reduction of the 
ionization balance in the stellar wind that emits the wideband 
Ly$\alpha^{\mathrm{sw}}$ radiation (Fig. 2c).

\begin{figure} %%Figure 6
\resizebox{8.8cm}{!}{\includegraphics{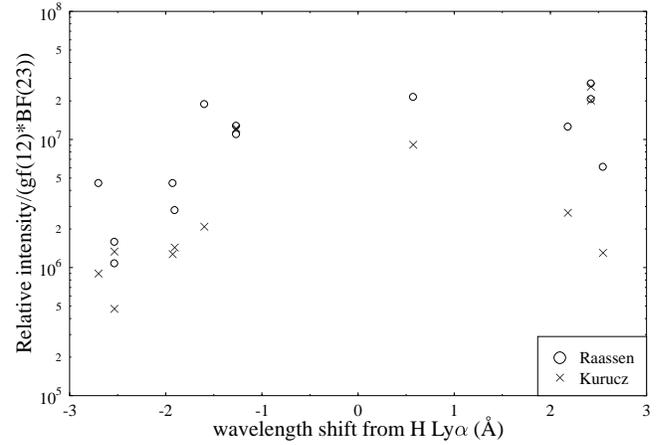}}
\caption{Indirect measurement of the shape of the HLy$\alpha$ profile at 1215 Å. The intensity at a certain wavelength is derived from the strength of a fluorescence line, adjusted for the strength of the excitation channel and 
the branching fraction of the actual fluorescence line. Two different sets of atomic data were used: (x) from \citet{K01} and 
(o) from \citet{R02}. For the points at +2.2 and +2.5, which correspond to 
upper levels that are known to be mixed, the atomic parameters have been adjusted.}
\end{figure}

The wavelength difference between the Ly$\alpha$ pumped \ion{Fe}{ii} lines,  
responsible for the occurrence of these 12 fluorescence lines, and Ly$\alpha$ itself
ranges between -2.6 and +2.5 Å, as shown in Fig. 6. The plot shows the intensity 
of Ly$\alpha$ as indirectly measured from the observed intensity in the fluorescence lines, 
which is corrected for the branching fraction of the fluorescence line and for 
the oscillator strength of the excitation channel. The lower levels of 
all the excitation channels belong to the ($^5$D)4s a$^4$D $LS$ term, and no 
correction for different populations in the different fine-structure levels has been 
applied. Two sets of theoretical atomic data for Fe II, \citet{K01} and \citet{R02},
have been used to derive the Ly$\alpha$ intensities from the fluorescence lines. The
resulting intensities are presented in Fig. 6 by crosses using atomic data from \citet{K01} 
and open circles using data from \citet{R02}. The differences reflect the difficulties
in calculating atomic parameters for complex spectra like Fe II. The two upper levels 
responsible for the excitation lines at 2.2 and 2.5 \AA\ in Fig. 6 are known to be heavily 
mixed making the corresponding calculated $f$-values very uncertain. 
Based on laboratory intensities we have for either of the two levels divided the total 
line strengths in two equal parts both for the excitation channels and the fluorescence lines.
This means that the excitation channels to the two upper levels are equally efficient, and
that the resulting fluorescence lines to a given lower state has the same strength. The 
Ly$\alpha$ pumped Fe II states represented in Fig. 6 are excited by various parts of the 
wide Ly$\alpha^{\mathrm{sw}}$ profile associated with the stellar wind (Fig. 2c). It 
should be emphasized that the width of the spectral interval ($\pm$2.5 
Å) corresponds to a terminal stellar wind velocity of $\pm$625 km/s relative to the 
the motion of the Weigelt blobs.

\begin{figure*} %%Figure 7
\sidecaption
\resizebox{12cm}{!}{\rotatebox{0}{\includegraphics{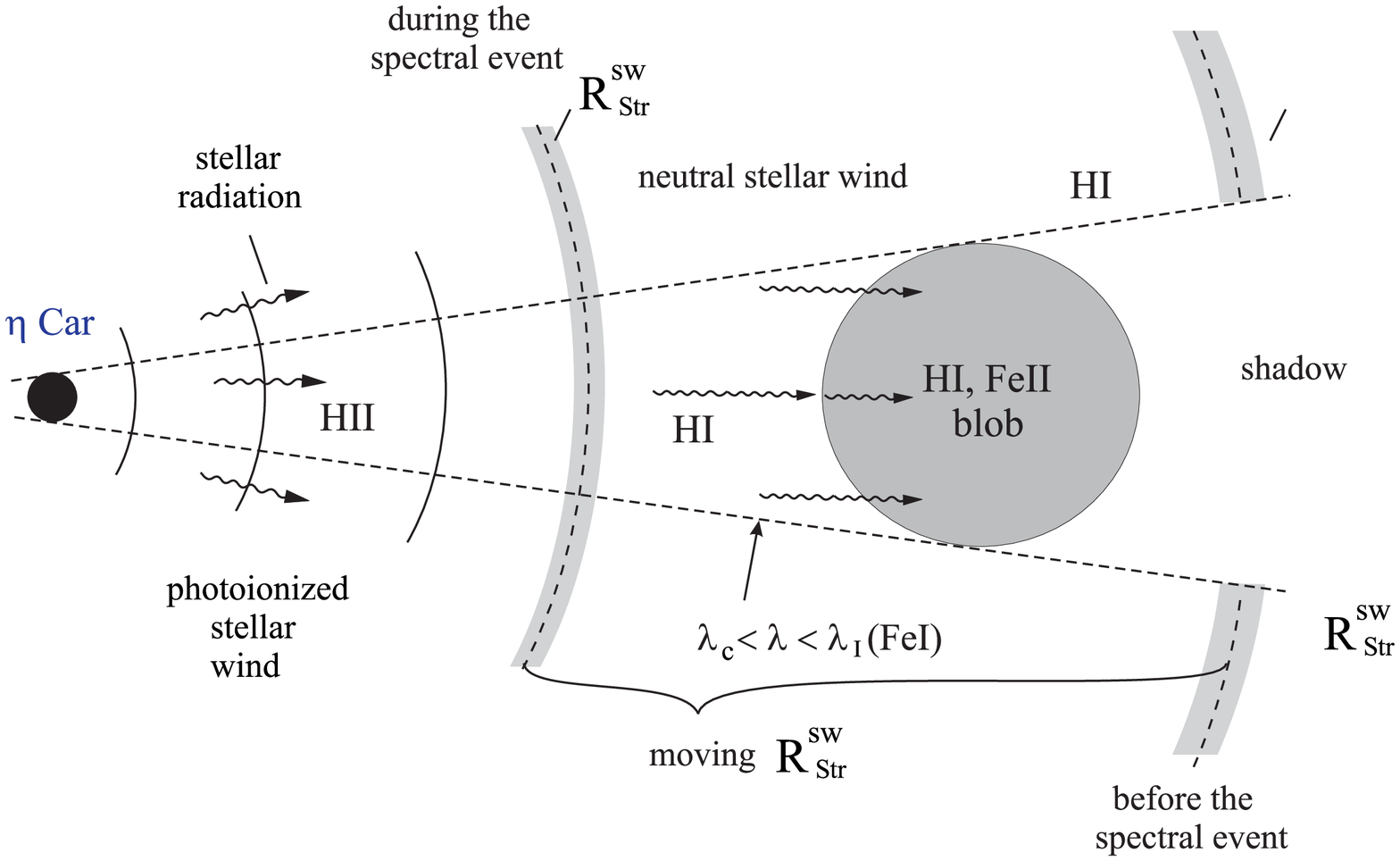}}}  %%*[97,106][788,453]
\caption{Sketch of the motion of the Strömgren boundary in the stellar wind toward the central source in Eta Carinae during the spectral event.}
\end{figure*}

During the spectral event 
there is a reduction in the intensity of the 
Ly$_{\mathrm{c}}$ radiation that ionizes both the blobs and the stellar 
wind (Fig. 1). As a result, the blobs become neutral with no 
frontal HII zone, and the Strömgren boundary in the stellar wind, 
$R_{\mathrm{Str}}^{\mathrm{sw}}$, shifts toward the central source, as shown 
in Fig. 7. A somewhat faster reduction of the intensity of some of the 
$<$\ion{Fe}{ii}> fluorescence lines would be a confirmation of such a 
shift of the Strömgren boundary for the stellar wind, because their 
excitation requires the red-shifted part of the Ly$\alpha^{\mathrm{sw}}$ profile,
which originates from the back side of the Weigelt blobs.

To illustrate the time dependence of the intensity reduction in the fluorescence lines for various
wavelength shifts of the Ly$\alpha$ radiation we have in Fig. 5b made normalized intensity reduction curves 
for the twelve $<$\ion{Fe}{ii}$>$ fluorescence lines. The relative intensity change with time 
for each  $<$\ion{Fe}{ii}$>$ line shows that those lines, whose excitation requires the 
red-shifted part of Ly$\alpha^{\mathrm{sw}}$, tend to 
decay 
a few %%approximately 5-6 
days sooner than the $<$\ion{Fe}{ii}$>$ lines 
excited by the blue-shifted part. This fact is consistent with %% supports 
our model proposing that 
the pumped \ion{Fe}{ii} levels are mainly excited by the 
Ly$\alpha^{\mathrm{sw}}$ radiation from the stellar wind, in which the 
Strömgren boundary is moving during the time of the spectral event 
\citep[see also ][]{JL04d}. %%Johansson and Letokhov, 2004b).

\begin{figure} %%Figure 8
\resizebox{8.8cm}{!}{\includegraphics{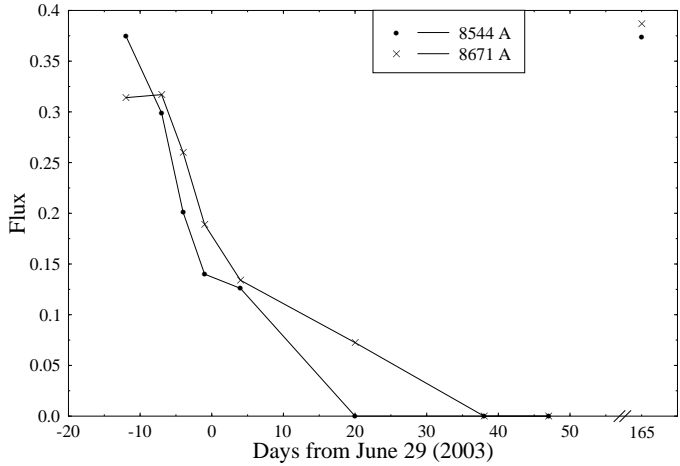}}
\caption{Time behaviour of \ion{Fe}{ii} laser lines during the spectroscopic event. The lines are 4s*$\rightarrow$4p transitions (B in Figure 3). Note the different flux scale compared to 5a.}
\end{figure}

Variations of the intensity of two spectral 
lines corresponding to 4s*$\rightarrow$4p transitions (transitions B in 
Fig. 3) are also observed. In this case the fluorescent decay is from 
the pseudo-metastable states 4s*, which have long lifetimes in spite of their
potential decay in $LS$-allowed transitions. These two fluorescence lines 
should exhibit a strong stimulated emission of radiation, since an 
inverted population is formed between the levels 4s* and 4p \citep{JL02a,JL03,JL04d}.
%%(Johansson and Letokhov, 2002; 2003; 2004b). 
The intensity variation character of 
these laser lines is similar to the behavior of the variation of the 
fluorescent 5p$\rightarrow$5s transitions (transition A 
in Fig. 3) featuring neither population inversion, nor stimulated emission 
effects. This can be explained by the fact that the emission intensity in 
transitions A and B is limited by the rate of pumping by the Ly$\alpha$ 
radiation, why it follows the pattern of the intensity variation of 
the Ly$\alpha$  pumping line. The only difference is that the intensity 
of the laser lines in the slow transition B is comparable with that 
of the fluorescence lines in the fast allowed transition A, just 
because of stimulated emission of radiation under saturated 
amplification conditions \citep{JL04d}. %%(Johansson and Letokhov, 2004).

\section{Intensity Variations of Forbidden \ion{Fe}{ii} Lines from Low-Lying Metastable States}

\begin{figure} %%Figure 9a,b
\resizebox{8.8cm}{!}{\includegraphics{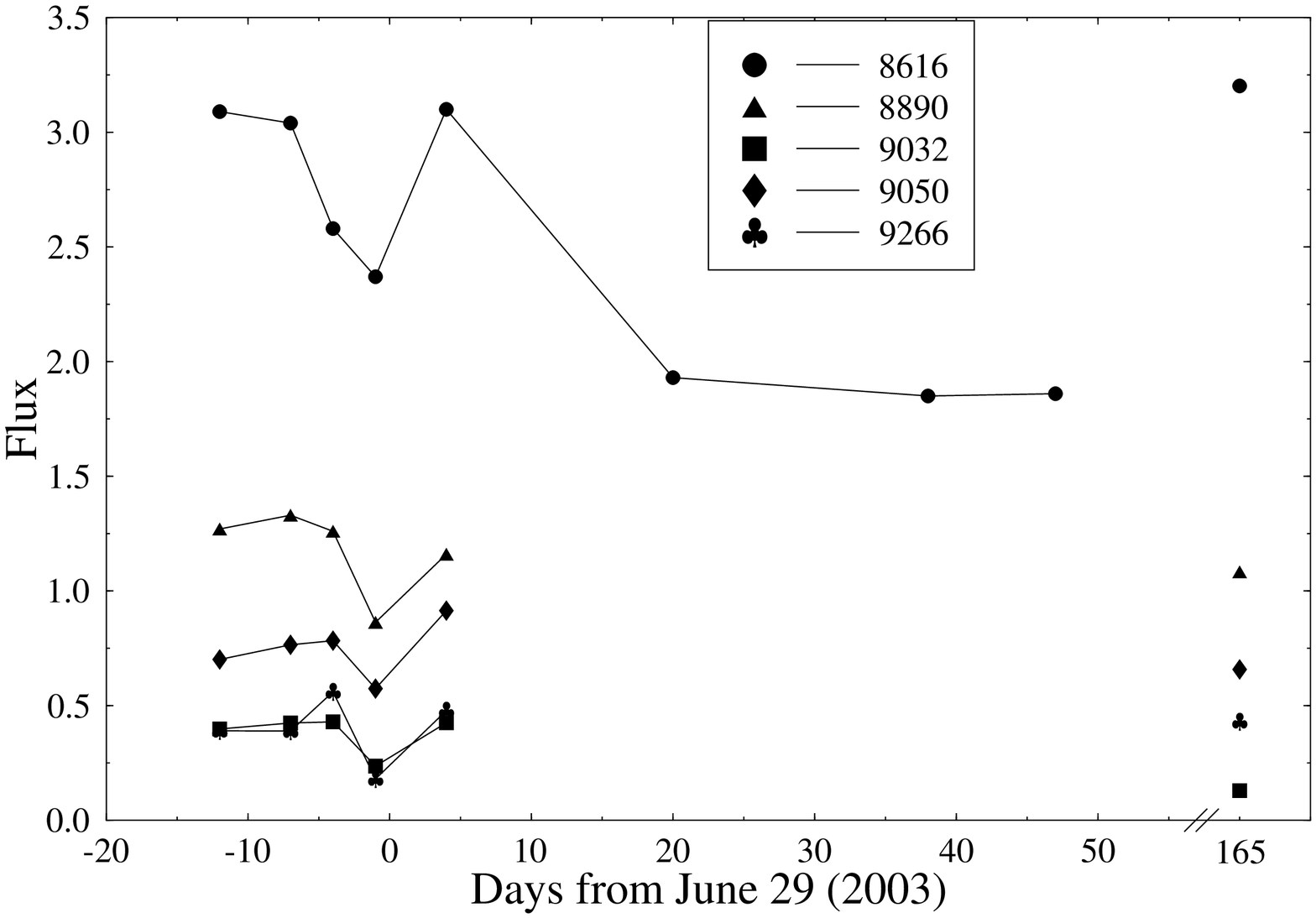}}
\resizebox{8.8cm}{!}{\includegraphics{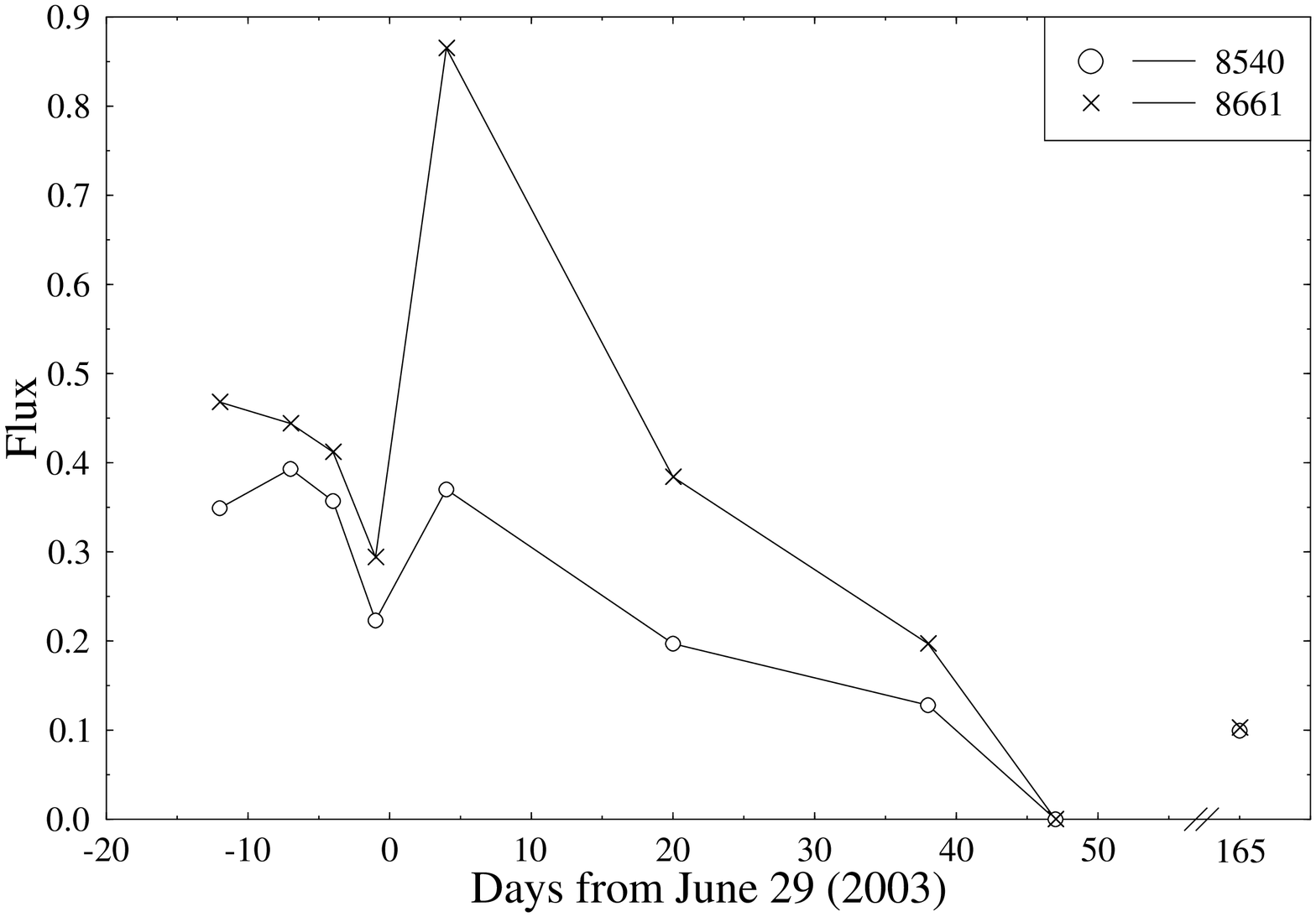}}
\caption{(a) Time behaviour of forbidden $[$\ion{Fe}{ii}$]$ lines during the spectroscopic event. The 
lines belong to the multiplet 3d$^7$~a$^4$F~-~3d$^7$~a$^4$P (12F). Only the 8616~\AA\ line was observed 
during the period after the event. Compare with the time behaviour of the fluorescent lines in Figure 4. 
(b) Same as 9a, but for CaII 3d\,$^2$D-4p\,$^2$P.}
\end{figure}

We have also studied intensity variation curves for five forbidden [\ion{Fe}{ii}] 
lines due to transitions between low-lying metastable states, as illustrated in Fig. 9. 
Their time behavior differs substantially from that of the $<$\ion{Fe}{ii}$>$ fluorescence 
lines associated with transitions between high-lying energy levels with short lifetimes.
The shape of the curves in Fig.~9 can be explained by 
the fact that the contribution to the population of the metastable states comes 
from two excitation mechanisms. One is the radiative decay of the high-lying states excited 
by Ly$\alpha^{\mathrm{sw}}$. The efficiency of this channel changes during the period of a 
spectral event, because it depends on the Ly$\alpha^{\mathrm{sw}}$ radiation 
and, in the final analysis, on the intensity of the stellar wind. The other 
mechanism is the collisional electronic excitation. This channel depends on the 
concentration of electrons in the HI zone. Since this channel does not decay 
during the period from 22 June to 15 August or longer (50-100 days), the 
electron recombination rate $W_{\mathrm{rec}} < 10^{-7} \mathrm{s}^{-1}$ and, 
according to Eq. (4), $n_{\mathrm{e}} < 3 \cdot 10^4 \mbox{ cm}^{-3}$. This 
agrees well with the electron concentration in the HI zone, governed by the 
photoionization of the Fe atoms, i.e., $n_{\mathrm{e}} \simeq 
N(\mathrm{\ion{Fe}{ii}}) \simeq 10^3-10^4 \mbox{ cm}^{-3}$.

An amazing fact is the growth of the intensity of the forbidden 
[\ion{Fe}{ii}] lines in the first half of July, when the spectral event 
is at its minimum. This can, in principle, be explained by a temporary 
increase in the electron density $n_{\mathrm{e}}$ for unknown reasons. 
Moreover, this possibility is probably supported by the fact that the same 
temporary intensity increase is exhibited by two Ca II lines. These are also due to 
transitions between relatively low-lying levels that can be excited 
by electrons. Naturally the absence in Ca II of low-lying metastable 
states and transitions coincident with the Ly$\alpha$ radiation makes 
the intensity behavior of the Ca II lines different from that of the 
\ion{Fe}{ii} lines.
 
\section{Conclusion}

The results extracted from the extensive ground-based spectral observations of  
H and \ion{Fe}{ii} lines formed in the Weigelt blobs of Eta~Carinae support 
and supplement the picture of the photophysical processes taking place in the 
near vicinity of the central source. The Weigelt blobs provide essentially a 
sensitive probe for these photoprocesses, especially during the course of the 
periodic and temporary reduction of the stellar radiation intensity during 
the spectral event. So far, we do not 
know the true source of the photoionizing radiation. Most likely it is the 
combined radiation from the primary and secondary stars in a binary system. One 
can suppose that during the spectral event the radiation from the hotter secondary 
star is temporarily screened by the primary star as seen from the Weigelt blobs. 
This leads to the reduction of the intensity of the ionizing radiation reaching the 
blobs for a period of the order of 100 days.

As a first result of the reduction of the ionizing radiation, the Strömgren boundary 
$R_{\mathrm{Str}}^{\mathrm{b}}$ disappears from the blobs, so that they 
become practically neutral. This manifests itself in the reduction of 
the intensity of the recombination Paschen lines from the HII zone of 
the blobs.

Secondly, the Strömgren boundary $R_{\mathrm{Str}}^{\mathrm{sw}}$ in the 
stellar wind moves closer to the central source. As a result, the  
red-shifted part of the $\mathrm{Ly}\alpha^{\mathrm{sw}}$ radiation 
disappears first and then the blue-shifted part. This is manifested in 
the behavior of the intensity reduction of the $<$\ion{Fe}{ii}$>$ lines. For 
the time being we can only speak of this tendency for the consecutive 
change in the intensity of the $<$\ion{Fe}{ii}$>$ lines, as we observe 
the total radiation in these lines from all the Weigelt blobs, which 
occupy somewhat different positions relative to the central source of Eta~Carinae.

The reduced $\mathrm{Ly}\alpha^{\mathrm{sw}}$ radiation results 
in some reduction in the intensities of the forbidden [\ion{Fe}{ii}] lines 
from low-lying metastable states in \ion{Fe}{ii}. This is due to the 
decrease in the population of these states resulting from the 
fluorescence decays of high-lying states. The remaining part of the intensity 
in the [\ion{Fe}{ii}] lines is due to excitation by collisions with 
electrons. The low concentration of electrons in the HI zone prevents 
the Fe$^+$ ions from recombination throughout the three months 
duration of the spectral event.

Finally, the intensity variation of the Paschen lines, the 
fluorescence <\ion{Fe}{ii}> lines, and the forbidden [\ion{Fe}{ii}] 
lines make it possible to estimate the concentration of electrons in the 
HI and HII zones of the Weigelt blobs. These qualitative estimates agree 
with the theoretical values calculated earlier.
\begin{acknowledgement}
The research project is supported by a grant (S.J.) from the Swedish National Space Board.
A.D.\ thank FAPESP and CNPq Brazilian agencies and to Lund University for support.
V.S.L.\ acknowledges the financial support through grants (S.J.) from the Wenner-Gren Foundations, as well as Lund Observatory for hospitality, and the Russian Foundation for Basic Research (Grant No.~03-02-16377). 
%%Wenner-Gren,Rymdstyrelsen,Damineli,Letokhov+Troitsk
\end{acknowledgement}

\bibliographystyle{aa}
\bibliography{e:/latex/bibtex/hartman}

\begin{thebibliography}{29}
\expandafter\ifx\csname natexlab\endcsname\relax\def\natexlab#1{#1}\fi

\bibitem[{{Bautista} \& {Pradhan}(1998)}]{BP98}
{Bautista}, M.~A. \& {Pradhan}, A.~K. 1998, \apj, 492, 650

\bibitem[{{Brown} {et~al.}(1979){Brown}, {Jordan}, \& {Wilson}}]{BJW79}
{Brown}, A., {Jordan}, C., \& {Wilson}, R. 1979, in The first year of IUE, ed.
  A.~{Willis}, 232

\bibitem[{{Damineli}(1996)}]{D96}
{Damineli}, A. 1996, ApJL, 460, L49

\bibitem[{{Damineli} {et~al.}(1997){Damineli}, {Conti}, \& {Lopes}}]{DCL97}
{Damineli}, A., {Conti}, P.~S., \& {Lopes}, D.~F. 1997, New Astronomy, 2, 107

\bibitem[{{Davidson}(2001)}]{D01}
{Davidson}, K. 2001, in Astronomical Society of the Pacific Conference Series,
  3--+

\bibitem[{{Davidson} {et~al.}(1997){Davidson}, {Ebbets}, {Johansson}, {Morse},
  \& {Hamann}}]{DEJ97}
{Davidson}, K., {Ebbets}, D., {Johansson}, S., {Morse}, J.~A., \& {Hamann},
  F.~W. 1997, AJ, 113, 335

\bibitem[{{Davidson} {et~al.}(1995){Davidson}, {Ebbets}, {Weigelt},
  {Humphreys}, {Hajian}, {Walborn}, \& {Rosa}}]{DEW95}
{Davidson}, K., {Ebbets}, D., {Weigelt}, G., {et~al.} 1995, AJ, 109, 1784

\bibitem[{{Davidson} \& {Humphreys}(1997)}]{DH97}
{Davidson}, K. \& {Humphreys}, R.~M. 1997, ARA\&A, 35, 1

\bibitem[{{Gull} {et~al.}(2001){Gull}, {Ishibashi}, {Davidson}, \&
  {Collins}}]{GID01}
{Gull}, T., {Ishibashi}, K., {Davidson}, K., \& {Collins}, N. 2001, in ASP
  Conf. Ser. 242: Eta Carinae and Other Mysterious Stars: The Hidden
  Opportunities of Emission Spectroscopy, 391

\bibitem[{{Gull} {et~al.}(1999){Gull}, {Ishibashi}, {Davidson}, \& {The Cycle 7
  STIS Go Team}}]{GID99}
{Gull}, T.~R., {Ishibashi}, K., {Davidson}, K., \& {The Cycle 7 STIS Go Team}.
  1999, in Eta Carinae at the Millennium, ASP Conf. Ser. 179. Edited by J.A.
  Morse, R.M. Humphreys, and A. Damineli., 144

\bibitem[{{Gull} {et~al.}(2005){Gull}, {Vieira}, {Bruhweiler}, {Nielsen},
  {Verner}, \& {Danks}}]{GVB05}
{Gull}, T.~R., {Vieira}, G., {Bruhweiler}, F., {et~al.} 2005, \apj, 620, 442

\bibitem[{{Johansson} \& {Hamann}(1993)}]{JH93}
{Johansson}, S. \& {Hamann}, F. 1993, Physica Scripta, T47, 157

\bibitem[{{Johansson} \& {Jordan}(1984)}]{JJ84}
{Johansson}, S. \& {Jordan}, C. 1984, MNRAS, 210, 239

\bibitem[{{Johansson} \& {Letokhov}(2001)}]{JL01e}
{Johansson}, S. \& {Letokhov}, V.~S. 2001, \aap, 378, 266

\bibitem[{{Johansson} \& {Letokhov}(2002)}]{JL02a}
{Johansson}, S. \& {Letokhov}, V.~S. 2002, Journal of Experimental and
  Theoretical Physics Letteres, 75, 495

\bibitem[{{Johansson} \& {Letokhov}(2003)}]{JL03}
{Johansson}, S. \& {Letokhov}, V.~S. 2003, Physical Review Letters, 90, 011101

\bibitem[{{Johansson} \& {Letokhov}(2004{\natexlab{a}})}]{JL04a}
{Johansson}, S. \& {Letokhov}, V.~S. 2004{\natexlab{a}}, Astronomy Letters, 30,
  58

\bibitem[{{Johansson} \& {Letokhov}(2004{\natexlab{b}})}]{JL04d}
{Johansson}, S. \& {Letokhov}, V.~S. 2004{\natexlab{b}}, \aap, 428, 497

\bibitem[{{Johansson} \& {Zethson}(1999)}]{JZ99}
{Johansson}, S. \& {Zethson}, T. 1999, in Astronomical Society of the Pacific
  Conference Series, 171

\bibitem[{{Klimov} {et~al.}(2002){Klimov}, {Johansson}, \& {Letokhov}}]{KJL02}
{Klimov}, V., {Johansson}, S., \& {Letokhov}, V.~S. 2002, \aap, 385, 313

\bibitem[{Kurucz(2001)}]{K01}
Kurucz, R. 2001, http://cfaku5.harvard.edu/atoms.html

\bibitem[{{Pittard} \& {Corcoran}(2002)}]{PC02}
{Pittard}, J.~M. \& {Corcoran}, M.~F. 2002, A\&A, 383, 636

\bibitem[{{Raassen}(2002)}]{R02}
{Raassen}, A. J.~J. 2002, http://www.science.uva.nl/pub/orth/iron/

\bibitem[{{Smith} {et~al.}(2004){Smith}, {Morse}, {Gull}, {Hillier}, {Gehrz},
  {Walborn}, {Bautista}, {Collins}, {Corcoran}, {Damineli}, {Hamann},
  {Hartman}, {Johansson}, {Stahl}, \& {Weis}}]{SMG04}
{Smith}, N., {Morse}, J.~A., {Gull}, T.~R., {et~al.} 2004, ApJ, 605, 405

\bibitem[{{Stahl} {et~al.}(2005){Stahl}, {Weis}, {Bomans}, {Davidson}, {Gull},
  \& {Humphreys}}]{SWB05}
{Stahl}, O., {Weis}, K., {Bomans}, D.~J., {et~al.} 2005, A\&A, in press

\bibitem[{{Steiner} \& {Damineli}(2004)}]{SD04}
{Steiner}, J.~E. \& {Damineli}, A. 2004, \apjl, 612, L133

\bibitem[{{van Boekel} {et~al.}(2003){van Boekel}, {Kervella}, {Sch{\" o}ller},
  {Herbst}, {Brandner}, {de Koter}, {Waters}, {Hillier}, {Paresce}, {Lenzen},
  \& {Lagrange}}]{BKS03}
{van Boekel}, R., {Kervella}, P., {Sch{\" o}ller}, M., {et~al.} 2003, \aap,
  410, L37

\bibitem[{{Verner} {et~al.}(2002){Verner}, {Gull}, {Bruhweiler}, {Johansson},
  {Ishibashi}, \& {Davidson}}]{VGB02}
{Verner}, E.~M., {Gull}, T.~R., {Bruhweiler}, F., {et~al.} 2002, ApJ, 581, 1154

\bibitem[{{Weis} {et~al.}(2005){Weis}, {Stahl}, {Bomans}, {Davidson}, {Gull},
  \& {Humphreys}}]{WSB05}
{Weis}, K., {Stahl}, O., {Bomans}, D.~J., {et~al.} 2005, \aj, 129, 1694

\end{thebibliography}

\end{document}